\documentclass[prl,twocolumn,showpacs,amsfonts,floatfix]{revtex4}
\usepackage{graphicx}
\usepackage{bm}
\newcommand{\n}{\nonumber}
\newcommand{\bn}{\begin{eqnarray}}
\newcommand{\en}{\end{eqnarray}}
\newcommand{\h}{\hspace}
\begin{document}
\title {Interference of a thermal Tonks gas on a ring}
\author{Kunal K. Das}
\email{kdas@optics.arizona.edu}
\author{M.D. Girardeau}
\email{girardeau@optics.arizona.edu}
\author{E.M. Wright}
\email{Ewan.Wright@optics.arizona.edu}
 \affiliation{Optical
Sciences Center and Department of Physics, University of Arizona,
Tucson, AZ 85721}
\date{\today}
\begin{abstract}
A nonzero temperature generalization of the Fermi-Bose mapping
theorem is used to study the exact quantum statistical dynamics of
a one-dimensional gas of impenetrable bosons on a ring. We
investigate the interference produced when an initially trapped
gas localized on one side of the ring is released, split via an
optical-dipole grating, and recombined on the other side of the
ring. Nonzero temperature is shown not to be a limitation to
obtaining high visibility fringes.
\end{abstract}
 \maketitle
A fundamental assumption at the heart of current proposals to
realize integrated atom sensors is that the guided atom
wavepackets will display interference phenomena when they are
split, propagated, and recombined. On the other hand it is well
known that the coherence properties of atomic gases are affected
by dimensionality, there being no true off-diagonal-long-range
order in less than three dimensions, and this raises the issue of
interference in restricted geometries. Motivated by recent
theoretical arguments \cite{MooreMeystre,KetterleInouye}
demonstrating that several stimulated processes for matter waves
such as four-wave mixing, superradiance, and matter-wave
amplification can be achieved in degenerate fermion gases as well
as in Bose-condensed gases, we have recently shown
\cite{RingTonks,X-Tonks} that a one-dimensional gas of hard-core
bosons, or Tonks gas, at zero temperature can exhibit high
visibility interference fringes. These results suggest that the
same mechanisms might be capable of overcoming the weakening of
interference due to thermal excitation. The quantum Tonks gas is
realized \cite{Ols98,PetShlWal00} in a regime essentially opposite
from that required for BEC, namely, the regime of low temperatures
and densities and large positive scattering lengths where the
transverse mode becomes frozen and the many-body Schr\"{o}dinger
dynamics becomes exactly soluble via a Fermi-Bose mapping theorem
\cite{Gir60,Gir65,RojCohBer99,GirWri00a,GirWri00b,GirWriTri01}. We
shall extend that theorem so as to obtain exact results for a
model atom interferometer using a thermal Tonks gas.  In
particular, we apply our results to the interference produced when
an initially trapped gas localized on one side of a ring is
released, split via an optical-dipole grating, and recombined on
the other side of the ring. Such a study is currently of relevance
due to experimental efforts
\cite{HinBosHug98,Schmied98,KeyHugRooSauHin00,Mul99,Thy99a,Dek00,HanReiHom01,
Ott01} to fabricate atomic waveguides for matter wave
interferometers.

\textit{Model:} The model consists of a 1D gas of $N_0$ hard core
bosonic atoms on a ring. This situation can be realized physically
using a toroidal trap of high aspect ratio $R=2L/\ell_{0}$ where
$2L$ is the torus circumference and $\ell_{0}$ the transverse
oscillator length $\ell_{0}=\sqrt{\hbar/m\omega_0}$ with
$\omega_0$ the frequency of transverse oscillations, assumed to be
harmonic. The longitudinal (circumferential) motion can be
described by a 1D coordinate $x$ along the ring with periodic
boundary conditions. The pure-state quantum dynamics of the system
is described by the time-dependent many-body Schr\"{o}dinger
equation (TDMBSE) $i\hbar\partial\Psi_B/\partial t=\hat H\Psi_B$
with Hamiltonian
\begin{equation}\label{eq1}
\hat{H}=-\frac{\hbar^2}{2m}\sum_{j=1}^{N}\frac{\partial^2}
{\partial x_{j}^{2}} +V(x_{1},\cdots,x_{N};t)  .
\end{equation}
Here $x_j$ is the 1D position of the $j^{\it th}$ particle,
$\psi_B(x_{1},\cdots,x_{N};t)$ is the N-particle wave function
with periodic boundary conditions,
\begin{equation}
\Psi_B(x_1\cdots x_j\pm 2L\cdots x_N;t) = \Psi_B(x_1\cdots
x_j\cdots x_N;t) ,
\end{equation}
which is also symmetric under exchange of any two particle
coordinates, as is the many-body potential $V$. We consider the
case of impenetrable two-particle interactions, the so-called
Tonks-gas regime \cite{Ols98,PetShlWal00}, and this is
conveniently treated as a constraint on allowed wave functions:
\begin{equation}\label{eq2}
\Psi_B=0\quad\text{if}\quad x_{j}=x_{k}\quad,\quad 1\le j<k\le N ,
\end{equation}
rather than as an infinite contribution to $V$, which then
consists of all other (finite) interactions and external
potentials.

We assume preparation of the system such that its initial state at
time $t=0$ is expressed by a statistical density operator
$\hat{\rho}_{0}=\sum_{\alpha}w_{\alpha}|\Psi_{B\alpha}(0)\rangle
\langle\Psi_{B\alpha}(0)|$ where the $|\Psi_{B\alpha}(0)\rangle$
are a complete set of orthonormal many-boson states with label
$\alpha$ that stands for a  set of quantum numbers, and the
$w_\alpha$ are nonnegative statistical weights summing to unity.
Then the statistical average of any observable $\hat{O}$ at any
later time $t>0$ is
\begin{equation}\label{evolve}
\langle\hat{O}(t)\rangle
=\sum_{\alpha}w_{\alpha}\langle\Psi_{B\alpha}(t)|\hat{O}|
\Psi_{B\alpha}(t)\rangle
\end{equation}
where the states $|\Psi_{B\alpha}(t)\rangle$ evolve according to
the TDMBSE  with Hamiltonian (\ref{eq1}).

\textit{Statistical Fermi-Bose mapping theorem:}
The Fermi-Bose mapping theorem for pure quantum states
\cite{Gir60,Gir65,RojCohBer99,GirWri00a,GirWri00b,GirWriTri01} is a mapping
$|\Psi_B\rangle=\hat{A}|\Psi_F\rangle$ from the Hilbert space of
one-dimensional many-fermion states $|\Psi_F\rangle$ to that of
one-dimensional many-boson states $|\Psi_B\rangle$, holding if the
interparticle interaction contains a hard core and the Hamiltonian
is of the form (\ref{eq1}). In Schr\"{o}dinger representation the
mapping operator $\hat{A}$ on $N$-particle states consists of multiplication
by the ``unit antisymmetric function''
\begin{equation}\label{eq3}
A(x_{1},\cdots,x_{N})=\prod_{1\le j<k\le N}\text{sgn}(x_{k}-x_{j}),
\end{equation}
where $\text{sgn}(x)$ is the algebraic sign of $x=x_{k}-x_{j}$,
i.e., it is +1(-1) if $x>0$($x<0$). Since $A$ is constant except
at nodes of the wavefunctions, the mapping converts antisymmetric
fermionic solutions $\Psi_{F}(x_{1},\cdots,x_{N};t)$ of the TDMBSE
into symmetric bosonic solutions $\Psi_{B}(x_{1},\cdots,x_{N};t)$
satisfying the same TDMBSE with the same Hamiltonian (\ref{eq1}),
and satisfying the same boundary conditions and constraint
(\ref{eq2}). In the Olshanii limit \cite{Ols98} (low density,
tightly confining atom waveguide, large scattering length) the
dynamics reduces to that of the impenetrable point Bose gas, which
is then satisfied trivially ($\Psi_F$ vanishing when any
$x_{j}=x_{\ell}$) due to antisymmetry. If the potential in Eq.
(\ref{eq1}) is a sum of one-body external potentials,
$V(x_{1},\cdots,x_{N};t)=\sum_{i}^{N}V(x_{i},t)$, the solutions of
the fermion TDMBSE can be written as Slater determinants
\cite{Gir60,Gir65,RojCohBer99,GirWri00a,GirWri00b,GirWriTri01}
\bn\label{eq5}
\Psi_{F\alpha}(x_{1},\cdots,x_{N};t)=\frac{1}{\sqrt{N!}}
\det_{i,j=1}^{N}\phi_{n_i}(x_{j},t) \en
where $n_{i}\in\{0,1,\cdots\}$ and the $\phi_{n_i}$ are
orthonormal solutions of the single particle time-dependent
Schr\"{o}dinger equation (TDSE) for potential $V(x,t)$.

The generalization to quantum statistical
evolution (\ref{evolve}) of many-boson systems is now straightforward:
\begin{equation}\label{evolve2}
\langle\hat{O}(t)\rangle_{B}
=\sum_{\alpha}w_{\alpha}\langle\Psi_{F\alpha}(t)|{\hat{A}}^{-1}\hat{O}\hat{A}|
\Psi_{F\alpha}(t)\rangle
\end{equation}
where ${\hat{A}}^{-1}$ is the inverse mapping from the Hilbert
space of bosonic states to that of fermionic states. This is most
useful if the observable $\hat{O}$ commutes with the mapping
operator $\hat{A}$ in which case the identity
${\hat{A}}^{-1}\hat{A}=1$ implies that
$\langle\hat{O}(t)\rangle_{B}=\langle\hat{O}(t)\rangle_{F}$,
reducing the statistical evolution of the $\hat{O}$ in the
impenetrable point Bose gas to that of the same observable in an
ensemble, with the same statistical weights, of ideal Fermi gas
states $|\Psi_{F\alpha}\rangle$ related to the corresponding
impenetrable point Bose gas states $|\Psi_{B\alpha}\rangle$ by the
pure-state mapping theorem
\cite{Gir60,Gir65,RojCohBer99,GirWri00a,GirWri00b,GirWriTri01}.

\textit{Single particle density:} Interference fringes arising
from spatial density modulation are due to first-order coherence
manifested in the single-particle density $\rho(x,t)$; the
corresponding operator commutes with $\hat{A}$ since it depends
only on particle positions. It follows that the density profiles
for the Fermi and Bose problems are identical, as can also be seen
directly from the fact that $A^2(x_{1},\cdots,x_{N})=1$ and hence
$|\Psi_B|^2=|\Psi_F|^2$.

The pure state mapping theorem ensures that corresponding Bose and
Fermi states have the same particle number $N$ and energy levels
$E_{\nu}(N)$,  so that both systems have the same chemical
potential $\mu$, and therefore corresponding states labeled by
$\alpha\equiv\{N,E_{\nu}(N)\}$ will have the same grand canonical
statistical weights $w_{\alpha}=e^{(\mu N- E_{\nu}(N))}$
\cite{YangYang}. Thus we will describe the gas of impenetrable
bosons by a grand canonical ensemble of ideal Fermi gas states
$\Psi_{F\alpha}(x_{1},\cdots,x_{N};t)$, assuming that the mean
number of atoms $N_0$ is sufficiently large to justify a grand
canonical picture. At time $t=0$ we consider a basis of energy
eigenstates in the external potential $V(x,0)$. The thermal
average of the single particle density of the system is
\bn\label{gcensemble} \langle \rho(x,t)
\rangle=\frac{\sum_{N=0}^{\infty}\sum_{\nu}\rho_{N,E_\nu}(x,t)e^{(\mu
N-E_{\nu}(N))}}{\sum_{N=0}^{\infty}\sum_{\nu}e^{(\mu N-
E_{\nu}(N))}},\en
where the denominator is the grand partition function and
$\rho_{N,E_\nu}(x,t)=\sum_{i=1}^{N}|\phi_{n_{i}}(x,t)|^{2}$ is the
spatial density of a $N$-particle energy eigenstate of a free
Fermi gas , its energy $E_{\nu}(N)=\sum_{i=1}^{N}\epsilon_{n_i}$
being the sum of the energies $\epsilon_{n_i}$ of the single
particle states $\phi_{n_{i}}(x,t)$. Some manipulation of the sums
leads to the familiar form
\bn\label{density}
\langle\rho(x,t)\rangle=\sum_{n=0}^{\infty}f_{n}|\phi_n(x,t)|^2,\en
where $f_{n}=1/(e^{(\epsilon_{n}-\mu)/k_BT}+1)$ are Fermi-Dirac
occupation numbers. The chemical potential $\mu$ is determined by
$\sum_{n=0}^{\infty}f_{n}=N_{0}$. The density at a later time
$t>0$ is obtained from Eq.~(\ref{density}) after propagating each
orbital $\phi_{n}$ separately according to the TDSE for potential
$V(x,t)$.

\begin{figure}
\includegraphics[width=7cm,angle=0]{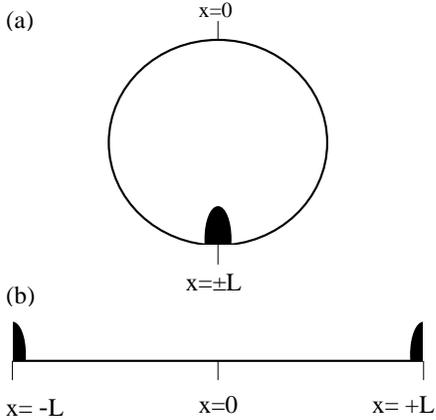}
\vspace{-3.2cm} \caption{(a) Initial density of $N$ hard core
bosons trapped on a ring of circumference $2L$. (b) By unfolding
the ring we describe the system using a 1D coordinate $x\in
[-L,L]$. The coincident point $x=-L\equiv L$ is chosen at the
center of the initial trapped gas.}
\label{Fig1.thermal-Tonks}\vspace{-5mm}
\end{figure}

\textit{Initial condition:} We assume that the ring has been
loaded (say, by optical tweezers \cite{Gus02}) with $N_0$ atoms at
temperature $T$, and for $t\le 0$ they are confined to a narrow
segment of the ring by a trapping potential $V(x,t\le 0)$ assumed
harmonic, with natural frequency $\omega$, over the spatial extent
of the initial trapped gas. The grand canonical ensemble in
Eqs.~(\ref{gcensemble}) and (\ref{density}) describing the state
of the atoms at $t=0$ is therefore characterized by the single
particle energies $\epsilon_{n}=(n+\frac{1}{2})\hbar\omega$. The
basic configuration is shown schematically in
Fig.~\ref{Fig1.thermal-Tonks}(a).

In order to discuss the time-evolution of the system, we designate
the normal modes of a class of 1D harmonic oscillators by
$u_{n}(x\!-\bar{x},w)$ where $\bar{x}$ are their mean positions
and the parameter $w$ relates their widths $x_{0}\sqrt{1+w^2}$ to
the width of the initial trap $x_{0}=\sqrt{\hbar/m\omega}$. Thus
the modes of the initial trap potential are
$u_{n}(x\!-\bar{x},w=0)\!$. We choose the circumferential
coordinates to have the trap center at $x=L\equiv -L$. On
unwrapping the ring about $x=0$ in
Fig.~\ref{Fig1.thermal-Tonks}(b), the initial Hermite-Gaussian
orbitals are split into two parts at the ends of the fundamental
periodicity cell $[-L,L]$, and can be written as
\begin{equation}\label{phizero}
\phi_{n}(x,t=0)=u_{n}(x+L,0)+u_{n}(x-L,0)\quad .
\end{equation}
The asymptotic dependence of the harmonic oscillator modes
$u_{n}(x)\simeq \cos(\sqrt{2n}x/x_{0})$ suggests the definition of
a critical wavevector for a thermal distribution,
$\bar{k}=\sqrt{2\bar{n}}/x_{0}$, which would give a measure of the
characteristic rate of expansion of atoms released from the
initial longitudinal trap.  We take it to correspond to
$f_{\bar{n}}=f_{0}/2$, so that for temperatures large compared to
the Fermi temperature ($T_{F}$) the critical wavevector coincides
with the thermal wavevector $\bar k\approx
\sqrt{2k_BT/\hbar\omega}/x_0$, and for $T\rightarrow 0$ we get the
wavevector corresponding to the highest occupied mode
$\bar{k}\rightarrow \sqrt{2N_{0}}/x_{0}$. If the initially trapped
gas is allowed to expand freely along the ring, the critical
wave-vector gives a measure of the time to wrap around the ring
$t_{wrap}=L/(\hbar \bar{k}/m)$.

Some necessary conditions need to be satisfied in order that the
initial atom cloud be accurately represented by a Tonks gas
\cite{Ols98,PetShlWal00}. First, the longitudinal energy must be
small compared with the transverse excitation energy: at zero
temperature this requires $N\hbar\omega\ll \hbar\omega_0$ and at
finite temperatures $k_{B}T\ll\hbar\omega_0$. Second, for the
initial gas to be accurately described as an impenetrable gas of
bosons we require $\bar{k}|a_{1D}|\ll 1$, i.e. $k_{B}T\ll
m\omega_{0}^{2}a ^{2}/2$ \cite{Ols98}.

\textit{Optical-dipole grating:}
In order to produce interference from the initial trapped gas we
turn off the harmonic trap at $t=0$ and apply a temporally short
but intense spatially periodic potential of wavevector $k$. This
spatially periodic grating may be produced over the spatial extent
of the trapped gas, for example, using intersecting and
off-resonant pulsed laser beams to produce an optical-dipole
grating whose wavevector may be tuned by varying the intersection
angle. The applied periodic potential then produces
counter-propagating scattered atomic waves, or daughter waves,
from the initial gas, or mother, with momenta $\pm\hbar k$ and
these recombine on the opposite side of the ring at a time $t_{r}
= L/(\hbar k/m)=(\bar k/k)t_{wrap}$. For times $t<2t_{wrap}$ the
mother packet has not yet encircled the ring and the gas expands
freely as if on an infinite line.

Following Rojo {\it et. al.} \cite{RojCohBer99} we use a
delta-function approximation for the short pulse excitation of the
periodic grating at $t=0$, and implement the optical-dipole
grating using a phase-imprinting scheme according to which each
orbital just after the pulse at $t=0^{+}$ is changed to
\begin{eqnarray} \label{initcond}
\phi_{n}(x,0^{+})& =& e^{i\eta\cos[k(x-L)]}\phi_{n}(x,0)
\\ &\approx& \left [1+\frac{i\eta}{2}\left
(e^{ik(x-L)}+e^{-ik(x-L)}\right )\right ]\phi_{n}(x,0) ,\nonumber
\end{eqnarray}
where $\eta$ is the grating amplitude. The second line assumes
$\eta<1$ in which case the optical-dipole grating predominantly
produces two scattered copies of the initial mode travelling to
the left and right with wavevectors $\pm k$ in addition to the
initial parent mode $\phi_{n}(x,0)$.
\begin{figure}
\vspace{-1cm}
\includegraphics[width=\columnwidth,angle=0]{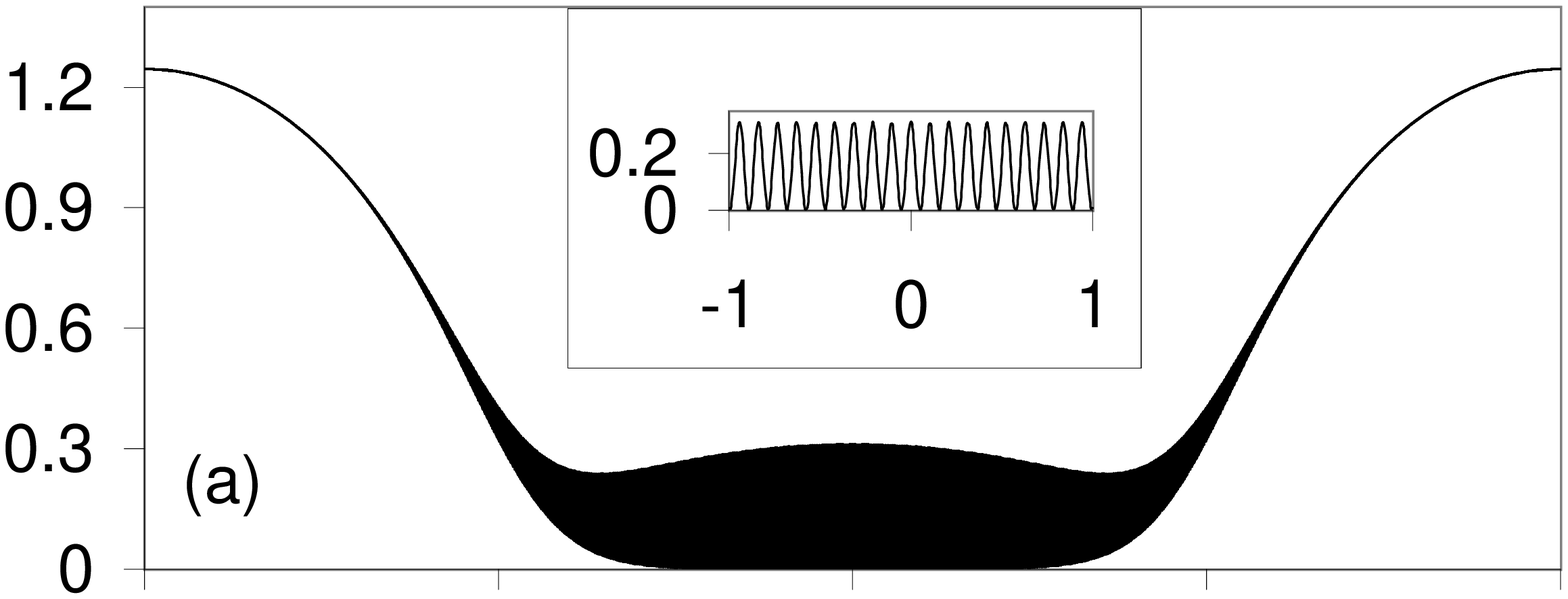}\vspace{-3.5cm}
\includegraphics[width=\columnwidth,angle=0]{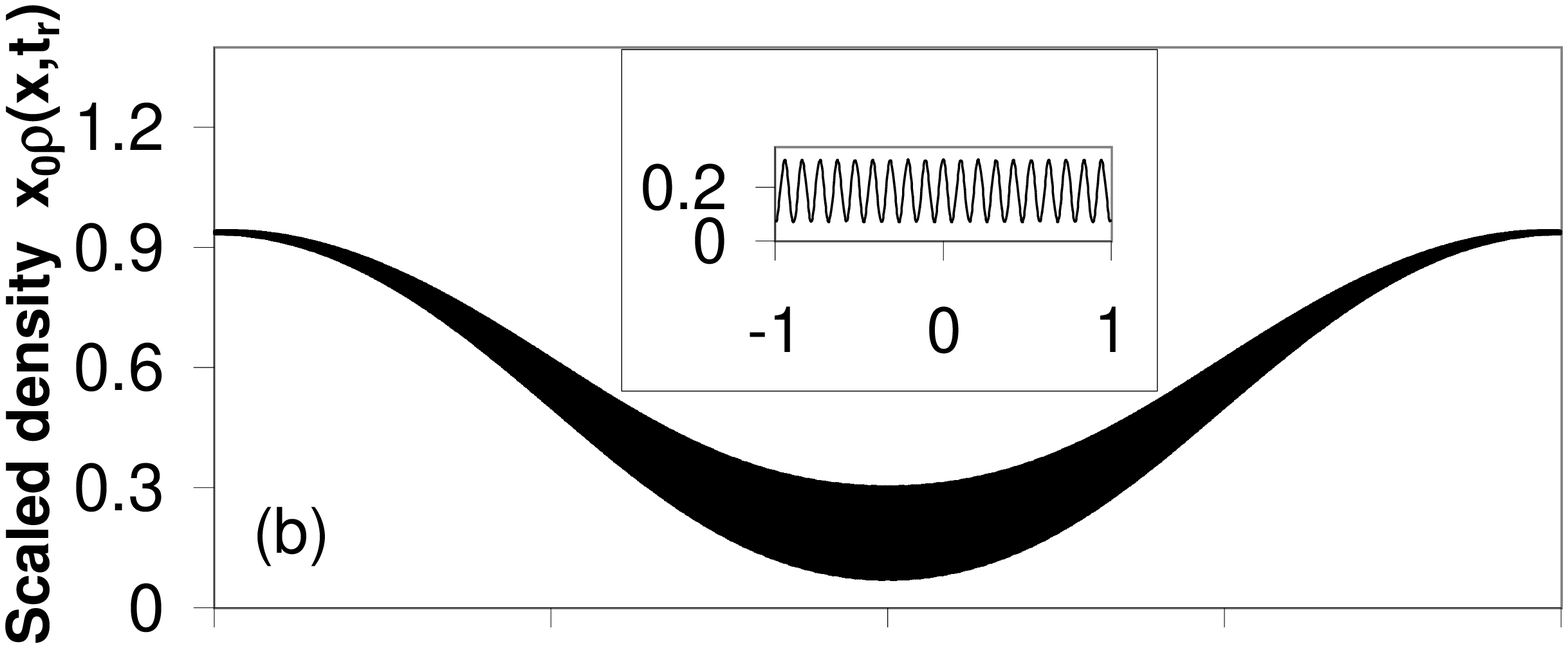}\vspace{-3.3cm}
\includegraphics[width=\columnwidth,angle=0]{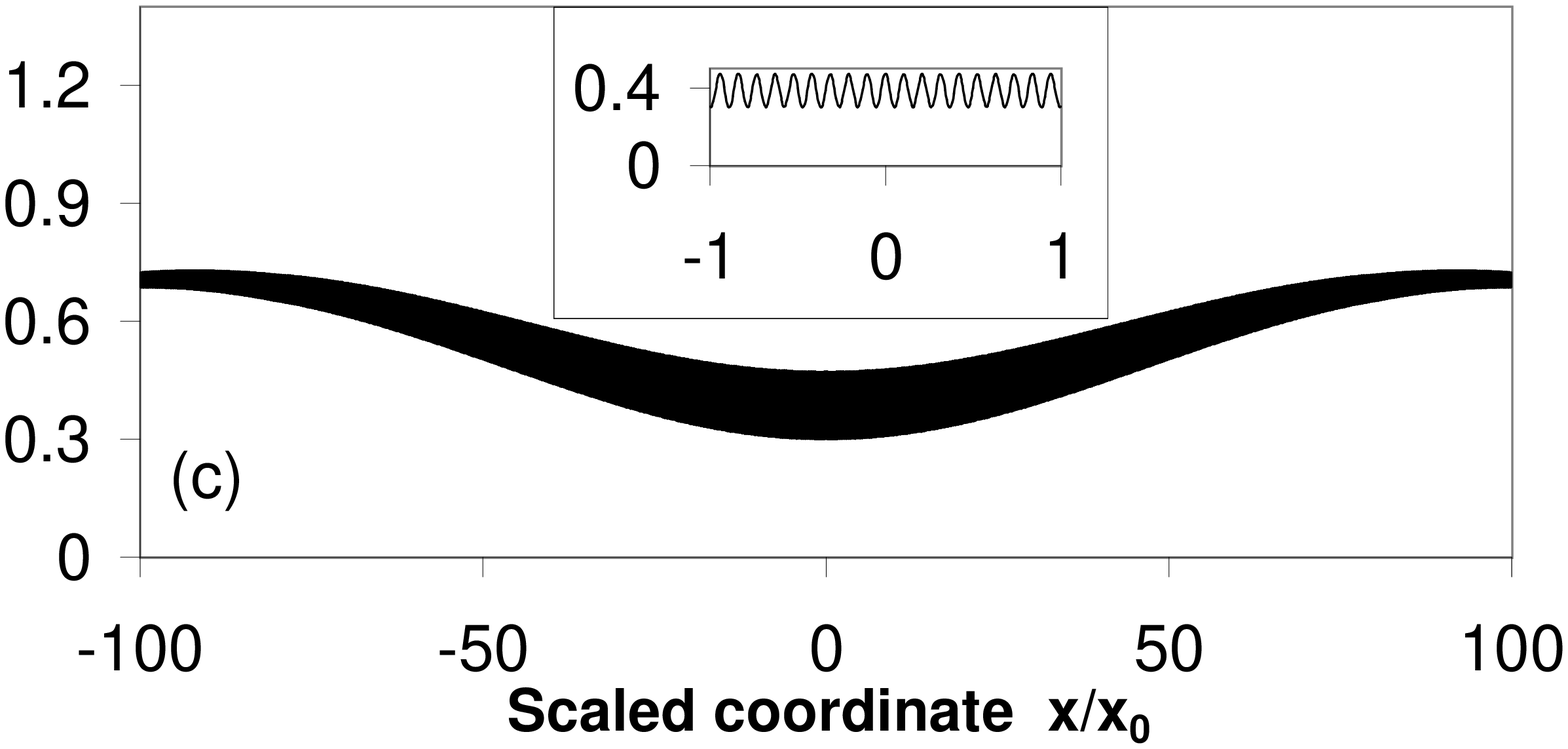}\vspace{-2.8cm}
 \caption{Scaled density profiles $x_{0}\rho(x,t_{r})$ for $N_{0}$=100 atoms,
 $\eta=0.5$, $L/x_{0}=100$ grating wavevector $x_{0}k=30$
shown at the time of recombination $t_r$ for temperatures (a)
$T=5$~nK, (b) $20$~nK and (c) $45$~nK. The insets show the details
of the fringes around the region of detection $x=0$. }
\label{Fig2.thermal-Tonks}\vspace{-5mm}
\end{figure}

\textit{Interference on a ring:} We wish to examine the presence
or absence of interference at the opposite side of the ring at
time $t=t_{r}$ and how this depends on the temperature $T$ and the
grating vector $k$. After the grating is applied the trapping
potential is turned off, $V(x,t>0)=0$, so the subsequent quantum
dynamics of the system is traced by freely propagating each
orbital $\phi_{n}(x,0^{+})$. Using the time evolved orbitals
$\phi_{n}(x,t_{r})$ in Eq.~(\ref{density}) we obtain the atom
density at time $t_r$ as a sum of three terms
$\rho(x,t_{r})=\rho_{m}(x,t_{r})+\rho_{d}(x,t_{r})+\rho_{md}(x,t_{r})$,
with
\bn\label{term1}
\rho_{m}(x,t_{r})=\sum_{n=0}^{\infty}f_{n}\left[u_{n}^{2}(x\!\!+\!\!L,\omega
t_{r})+ u_{n}^{2}(x\!\!-\!\!L,\omega
t_{r})\right.\h{6mm}\n\\\left.+2 u_{n}(x\!\!+\!\!L,\omega
t_{r})u_{n}(x\!\!-\!\!L,\omega t_{r}) \cos\left(\frac{2xL\omega
t_{r}}{x_{0}^{2}(1\!+\!\omega^{2}t_{r}^{2})}\right)\right]\n\\
\rho_{d}(x,t_{r})=\frac{\eta^{2}}{2}[1+\cos(2kx)]
\sum_{n=0}^{\infty}f_{n}u_{n}^{2}(x,\omega t_{r}).\h{1cm}\en
The first term $\rho_{m}$ is due to the expanding mother packet
overlapping with itself, $\rho_{d}$ is due to the overlap of the
daughters, and $\rho_{md}$ arises from the overlap of the mother
and the daughter packets. In general, for $t_{wrap}>t_r$, which is
desirable so that the mother packet does not wrap at the same time
as the daughters recombine, $\rho_{md}$ is small compared to the
other terms: This term is present in our simulation below but we
have refrained from reproducing the expression for it above.
\begin{figure}
\vspace{-1cm}
\includegraphics[width=\columnwidth,angle=0]{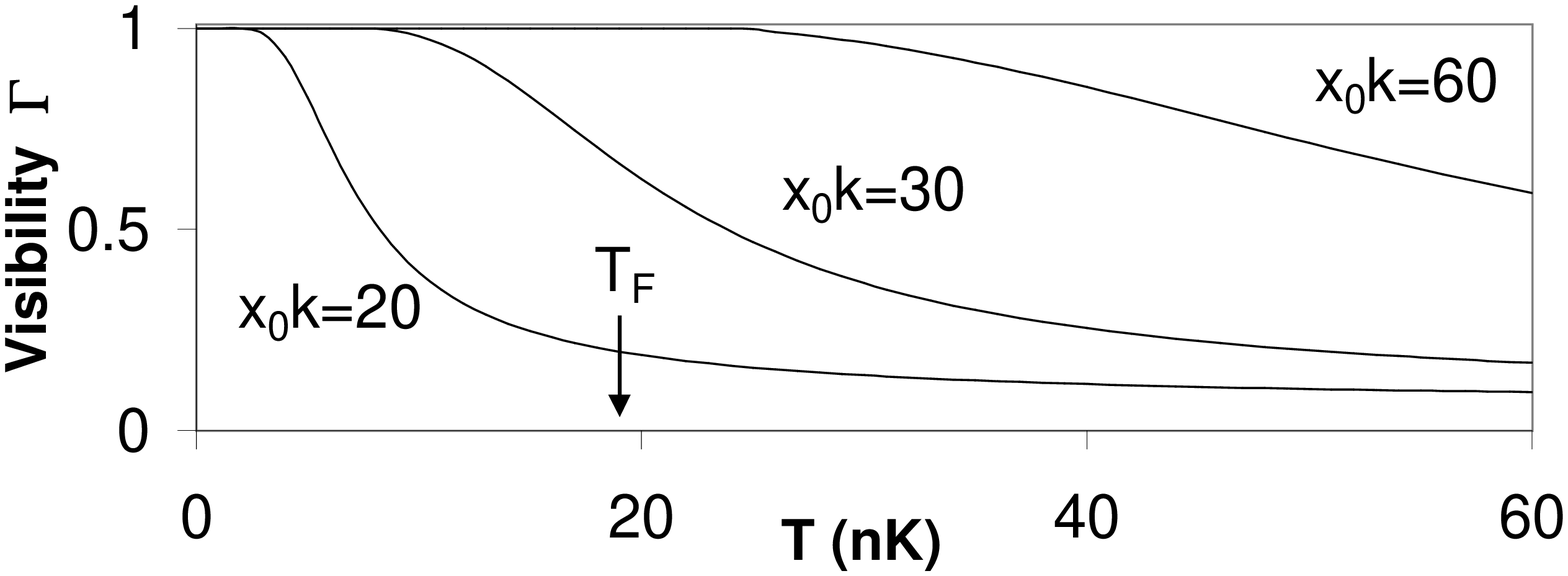}\vspace{-3cm}
 \caption{Estimate of the visibility in the region of detection, as a function
of temperature for three values of the optical-dipole grating
wavevector (a) $x_{0}k=20$, (b) $30$ and (c) $45$.}
\label{Fig3.thermal-Tonks}\vspace{-5mm}
\end{figure}

As a concrete example we consider $N_{0}=100$ sodium atoms on a
ring of circumference $2L=100 x_0$ released from an initial
harmonic trap of frequency $\omega=2\pi\times 4$ Hz. This
corresponds to an oscillator length of $x_{0}=10.4$ $\mu$m and
Fermi temperature $T_{F}=19.1$ nK. In general, we find that for
temperatures and applied wavevectors such that $k\ll\bar k$
minimal interference fringes result when the daughter packets
recombine at time $t_r$. Recalling that the wrap time for the
mother packet is given by $t_{wrap}=(k/\bar k)t_r$, then for
$k\ll\bar k$ the mother packet wraps around significantly by the
time the daughters recombine, so $\rho_{m}$ provides a pedestal
for interference fringes due to the overlap of the daughters
$\rho_{d}$, leading to reduced fringe visibility \cite{RingTonks}.
In contrast for $k\ge\bar k$ pronounced interference fringes can
appear when the daughter packets are recombined. This is
illustrated in Fig.~\ref{Fig2.thermal-Tonks} where we fix the
wavevector $k$ of the optical grating and vary the temperature or
alternatively $\bar k$. Unit visibility fringes are seen in
Fig.~\ref{Fig2.thermal-Tonks}(a) where $T=5$ nK, whereas reduced
visibility fringes appear in Figs.~\ref{Fig2.thermal-Tonks}(b,c)
where $T=20,45$ nK.

By noting that around $x=0$ the fringe pattern maximum of
$\rho(x,t_{r})$ is $\simeq\rho_{m}(0,t_{r})+\rho_{d}(0,t_{r})$,
and the minimum $\simeq~\rho_{m}(0,t_{r})$, we obtain a measure of
the fringe visibility as
\bn\Gamma=\frac{\rho_{d}(0,t_{r})}
{2\rho_{m}(0,t_{r})+\rho_{d}(0,t_{r})}.\en
This fringe visibility is plotted in
Fig~(\ref{Fig3.thermal-Tonks}) as a function of temperature for
three values of the grating vector $k$, with the clear and
physically obvious trend that higher temperatures require a higher
grating wavevector to obtain high visibility interference fringes.
Physically there are of course limitations: For sodium atoms, the
condition for a Tonks-gas is $T\ll (2\times 10^{-5}\times
\nu_{0})^{2}$~nK, for a transverse trapping frequency
$\omega_{0}=2\pi\nu_{0}$ Hz, and current experimental limits are
about $\nu_0\sim 2\times 10^{4}$ Hz \cite{Greiner}. For the sodium
yellow lines $k\sim 10^{7}$ m$^{-1}$, so for our parameters
$x_{0}k\sim 100$, in keeping with the parameters used above.

In conclusion, we have shown that even at nonzero temperature a
strongly interacting 1D gas of impenetrable bosons can show high
visibility  interference fringes on a ring. This remarkable result
is of importance for current schemes to realize integrated atom
interferometers in that it shows that neither many-body
interactions nor nonzero temperature are fundamental limitations,
in 1D at least. Our results followed from a generalization of the
zero temperature Fermi-Bose mapping which maps the strongly
interacting 1D boson problem to a 1D gas of free fermions, and it
is an open problem whether our conclusions can be extended to 2D
and 3D interferometers.
\begin{acknowledgments}
This work was
supported by Office of Naval Research Contract No.
N00014-99-1-0806 and by the U.S. Army Research Office.
\end{acknowledgments}
\end{document}